\documentclass[conference]{IEEEtran}
\usepackage{cite}
\usepackage{amsmath,amssymb,amsfonts}
\usepackage{algorithmic}
\usepackage{graphicx}
\usepackage{textcomp}
\usepackage{xcolor}
\def\BibTeX{{\rm B\kern-.05em{\sc i\kern-.025em b}\kern-.08em
    T\kern-.1667em\lower.7ex\hbox{E}\kern-.125emX}}
\begin{document}

\title{Lung Disease Detection with Vision Transformers: A Comparative Study of Machine Learning Methods}

\author{\IEEEauthorblockN{Baljinnyam Dayan}
\IEEEauthorblockA{\textit{Department of Computer Science and Information Technology} \\
\textit{National University of Mongolia}\\
\textit{Ulaanbaatar, Mongolia}\\
Baljinnyam.da@num.edu.mn}}

\maketitle

\begin{abstract}
Recent advancements in medical image analysis have predominantly relied on Convolutional Neural Networks (CNNs), achieving impressive performance in chest X-ray classification tasks, such as the 92\% AUC reported by AutoThorax-Net and the 88\% AUC achieved by ChexNet\cite{8099852,rajpurkar2017chexnetradiologistlevelpneumoniadetection, sze-to2021} in classifcation tasks. However, in the medical field, even small improvements in accuracy can have significant clinical implications. This study explores the application of Vision Transformers (ViT) \cite{vit_dosovitskiy2021imageworth16x16words,transformers_vaswani2023attentionneed}, a state-of-the-art architecture in machine learning, to chest X-ray analysis, aiming to push the boundaries of diagnostic accuracy.
I present a comparative analysis of two ViT-based approaches: one utilizing full chest X-ray images and another focusing on segmented lung regions. Experiments demonstrate that both methods surpass the performance of traditional CNN-based models, with the full-image ViT achieving up to 97.83\% accuracy and the lung-segmented ViT reaching 96.58\% accuracy in classifcation of diseases on three label and AUC of 94.54\% when label numbers are increased to eight. Notably, the full-image approach showed superior performance across all metrics, including precision, recall, F1 score, and AUC-ROC.
These findings suggest that Vision Transformers can effectively capture relevant features from chest X-rays without the need for explicit lung segmentation, potentially simplifying the preprocessing pipeline while maintaining high accuracy. This research contributes to the growing body of evidence supporting the efficacy of transformer-based architectures in medical image analysis and highlights their potential to enhance diagnostic precision in clinical settings.
\end{abstract}


\section{Introduction}
The early and accurate detection of lung diseases, such as pneumonia, COVID-19 and other diseases, is critical for reducing patient mortality and optimizing healthcare outcomes. Chest X-rays are one of the most widely used imaging modalities for diagnosing lung conditions due to their cost-effectiveness and accessibility. However, interpreting these images remains challenging, as pathological patterns can be subtle and complex. To address these challenges, automated methods based on machine learning have become a focus of research, offering the potential to assist radiologists by providing reliable diagnostic support \cite{litjens2017}.

In recent years, Convolutional Neural Networks (CNNs) have been the dominant architecture for medical image classification tasks, particularly in chest X-ray analysis. CNN-based models have demonstrated considerable success, achieving high accuracy on various lung disease detection tasks\cite{chexpert_Irvin_Rajpurkar_Ko_Yu_Ciurea-Ilcus_Chute_Marklund_Haghgoo_Ball_Shpanskaya_Seekins_Mong_Halabi_Sandberg_Jones_Larson_Langlotz_Patel_Lungren_Ng_2019}. Despite this success, further advancements are needed to improve diagnostic accuracy, as even marginal improvements can have significant clinical implications. Additionally, CNNs exhibit certain limitations, particularly in capturing long-range dependencies and global context within images. This shortfall may hinder their ability to detect subtle disease patterns, which are crucial for early diagnosis.

Vision Transformers (ViTs), a novel class of neural networks, have emerged as a powerful alternative to CNNs. Originally designed for natural image classification, ViTs utilize the self-attention mechanism to model global relationships across the entire image, making them particularly well-suited for tasks requiring a comprehensive understanding of image features\cite{vit_dosovitskiy2021imageworth16x16words}. In this study, I investigate the potential of ViTs to improve diagnostic accuracy in chest X-ray analysis.

The goal of this research is twofold: (1) to evaluate the performance of ViTs on full chest X-ray images, and (2) to explore whether focusing on lung-segmented regions can further enhance the accuracy of disease classification. By conducting a comparative analysis of these two approaches, I aim to determine which method better captures the relevant features necessary for detecting lung diseases.

\section{Related Work}

Early models like CheXNet and CheXpert leveraged CNNs to detect various lung diseases, achieving remarkable performance across large datasets \cite{chexpert_Irvin_Rajpurkar_Ko_Yu_Ciurea-Ilcus_Chute_Marklund_Haghgoo_Ball_Shpanskaya_Seekins_Mong_Halabi_Sandberg_Jones_Larson_Langlotz_Patel_Lungren_Ng_2019, rajpurkar2017chexnetradiologistlevelpneumoniadetection}. These models relied on the hierarchical feature extraction capability of CNNs, where local features were learned at lower layers and combined into more complex global patterns at deeper layers.

To enhance CNN performance, researchers introduced attention mechanisms that allowed the model to focus on more relevant parts of the image. Global attention strategies, like the ones introduced in models with self-attention layers, aimed to capture long-range dependencies across the entire image, while local attention was utilized to focus on critical regions such as lung areas. Hybrid approaches that combined CNNs with lung segmentation also improved model precision by narrowing the focus to specific regions of interest.

Despite the success of CNNs, their inherent limitation lies in their reliance on local feature extraction. This limitation prompted the exploration of Vision Transformers (ViTs), which use global self-attention mechanisms to capture broader context across images. 

\section{Method}
\subsection{Dataset}
This study utilizes two prominent datasets for chest X-ray image analysis: the NIH Chest X-ray dataset and the COVID-19 Image Data Collection. The NIH Chest X-ray dataset, provided by the National Institutes of Health (NIH), contains over 100,000 frontal-view X-ray images, covering 14 pulmonary conditions, including pneumonia, effusion, and atelectasis\cite{wang2017chestx}. The COVID-19 Image Data Collection, compiled by Joseph Paul Cohen, Paul Morrison, and Lan Dao, supplements this dataset with X-ray images of patients diagnosed with COVID-19\cite{cohen2020covid}. Together, these datasets form the foundation of this study’s exploration of lung disease classification using Vision Transformers (ViT).
\begin{figure}[h!]
    \centering
    \includegraphics[width=1\linewidth]{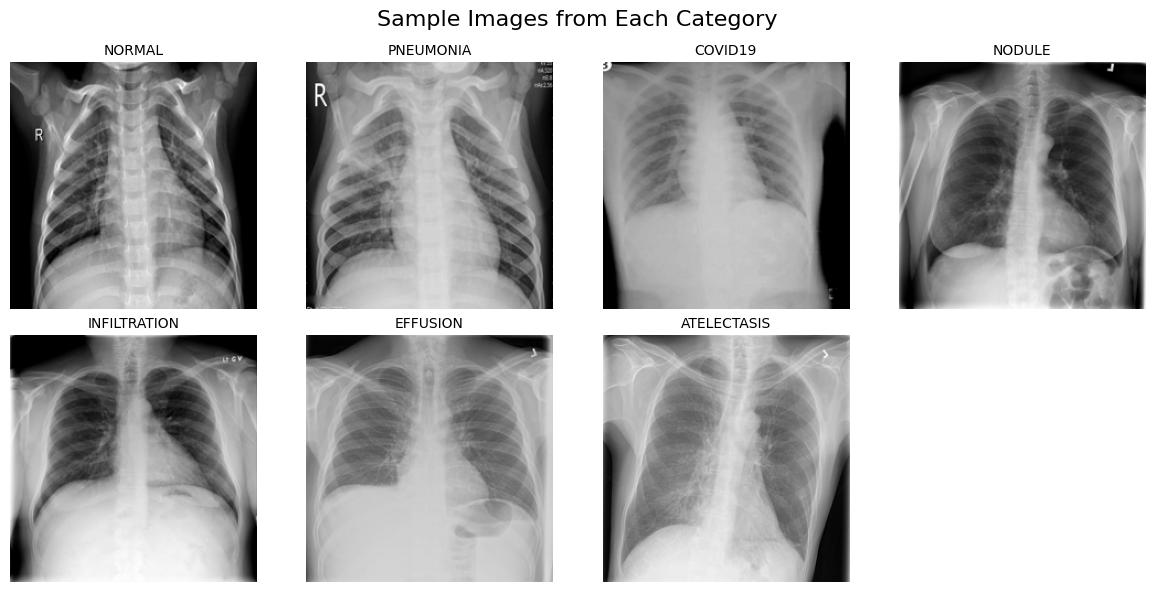}
    \caption{Sample pictures}
    \label{fig:enter-label}
\end{figure}
From these datasets, a subset was selected to ensure a balanced and representative sample of lung diseases. The training set includes a total of 12,897 images across seven categories, including 1,266 images for "Normal" cases, 1,598 for "Effusion," 2,000 for "Infiltration," 1,471 for "Nodule," 3,418 for "Pneumonia," 1,684 for "Atelectasis," and 460 for "COVID-19." Similarly, the test set contains 2,975 images, with 317 images in the "Normal" class, 399 for "Effusion," 500 for "Infiltration," 367 for "Nodule," 855 for "Pneumonia," 421 for "Atelectasis," and 116 for "COVID-19."
\begin{figure}[h!]
    \centering
    \includegraphics[width=1\linewidth]{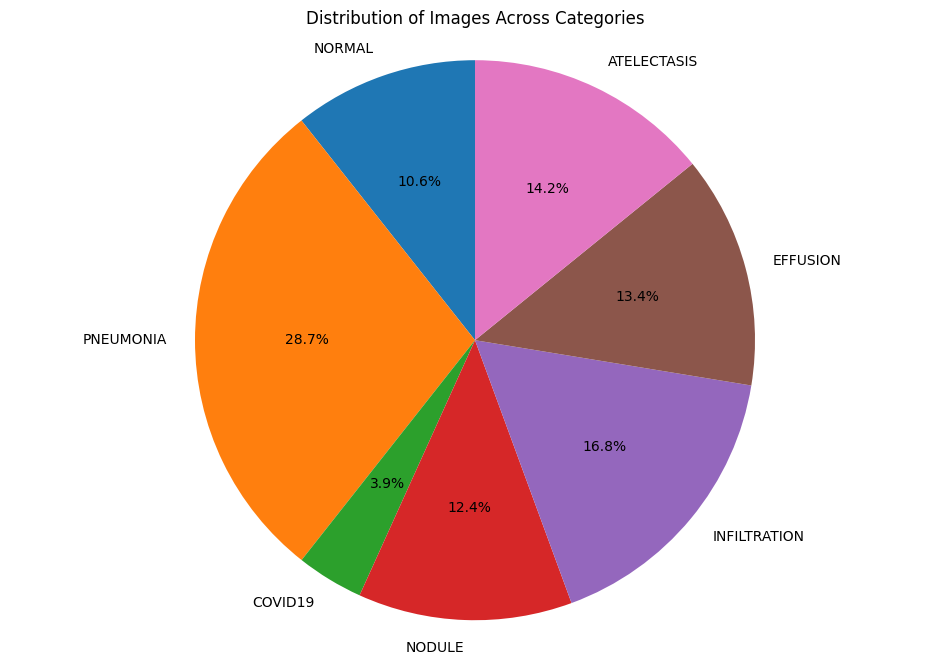}
    \caption{Distribution of Images}
    \label{fig:Dataset}
\end{figure}

In addition to these raw images, Segment Anything Model 2 (SAM2) was employed to segment the lung regions which was release July 29, 2024 from Meta AI, generating a masked dataset for a region-focused approach \cite{ravi2024sam2segmentimages}. This segmentation process allowed for an in-depth comparison between models trained on full chest X-rays and those trained on segmented lung areas, providing insights into the impact of localized versus global feature extraction in chest X-ray classification.

\subsection{Model Architecture}
For this study, I utilized the \textbf{google/vit-base-patch16-224-in21k} Vision Transformer (ViT) model as base architecture. The ViT architecture is based on a pure attention mechanism rather than the convolutional operations traditionally used in medical image analysis. It divides the input image into patches and processes them as sequences. Specifically, the model divides input images into patches of 16x16 pixels, which are then processed as a sequence.

These patches are embedded into a lower-dimensional space, followed by position embeddings to retain spatial information, and subsequently passed through multiple layers of self-attention. The self-attention mechanism is defined as:

\[
\text{Attention}(Q, K, V) = \text{softmax}\left(\frac{QK^\top}{\sqrt{d_k}}\right)V
\]

where \(Q\), \(K\), and \(V\) represent the query, key, and value matrices, and \(d_k\) is the dimension of the key vectors. This mechanism enables the model to capture both local and global features of the chest X-ray images.

Chosen ViT model was pre-trained on the ImageNet-21k dataset, providing a rich foundation of visual features. I fine-tuned this model for specific chest X-ray classification task. The model was fine-tuned with a custom classifier head consisting of a fully connected layer. For classification, a softmax layer was applied over the output logits to predict the probability distribution across the seven disease categories. I replaced the original classification head with a custom fully connected layer to output probabilities for seven disease categories.

The core components of the model are as follows:
\begin{itemize}
    \item \textbf{ViT Patch Embeddings}: The input images are divided into non-overlapping patches, each of size 16x16, which are then flattened and linearly projected into a higher-dimensional space (768 dimensions). The embeddings are computed using a patch embedding layer with 590,592 parameters.
    \item \textbf{ViT Encoder}: The sequence of image patches is fed into a transformer encoder, consisting of 12 layers of multi-head self-attention mechanisms, layer normalization, and MLP blocks. This structure enables the model to capture both local and global information from the image.
    \item \textbf{Classification Head}: After the encoding process, the class token is passed through a final fully connected layer, which outputs predictions for 7 lung disease categories (Normal, Pneumonia, COVID-19, Nodule, Infiltration, Effusion, and Atelectasis). The classification head includes a linear layer with 5,383 parameters.
\end{itemize}

The overall architecture comprises 597,511 parameters, all of which are trainable.

\subsection{Training Process}
Input chest X-ray images were preprocessed to match the model's requirements, including resizing to 224x224 pixels and normalizing pixel values.

The loss function for the optimization was the cross-entropy loss, denoted as:

\[
\mathcal{L}_{\text{CE}}(y, \hat{y}) = -\sum_{i=1}^{C} y_i \log(\hat{y}_i)
\]

where \(y\) is the true label, \(\hat{y}\) is the predicted probability, and \(C\) is the total number of classes. The optimization of model parameters was performed using the AdamW\cite{adamw_loshchilov2019decoupledweightdecayregularization} optimizer, with a learning rate of \(1 \times 10^{-4}\) and a weight decay of 0.01. The learning rate was scheduled using the cosine annealing method, where the learning rate is gradually reduced over each epoch according to the following formula:

\[
\eta_t = \eta_{\text{min}} + \frac{1}{2} (\eta_{\text{max}} - \eta_{\text{min}}) \left( 1 + \cos\left( \frac{t}{T_{\max}} \pi \right) \right)
\]

where \(\eta_t\) is the learning rate at epoch \(t\), \(\eta_{\text{min}}\) and \(\eta_{\text{max}}\) are the minimum and maximum learning rates, and \(T_{\max}\) is the total number of epochs.

The segmented dataset was used in parallel with the full-image dataset to evaluate the effectiveness of a localized approach in improving classification accuracy. I compared the model's ability to capture relevant features from both full and segmented X-ray images to assess its overall performance. This dual approach allowed us to investigate whether focusing on lung regions could enhance the model's diagnostic accuracy, results are shown in Figure~\ref{fig:enter-label}.
\begin{figure}[h!]
    \centering
    \includegraphics[width=1\linewidth]{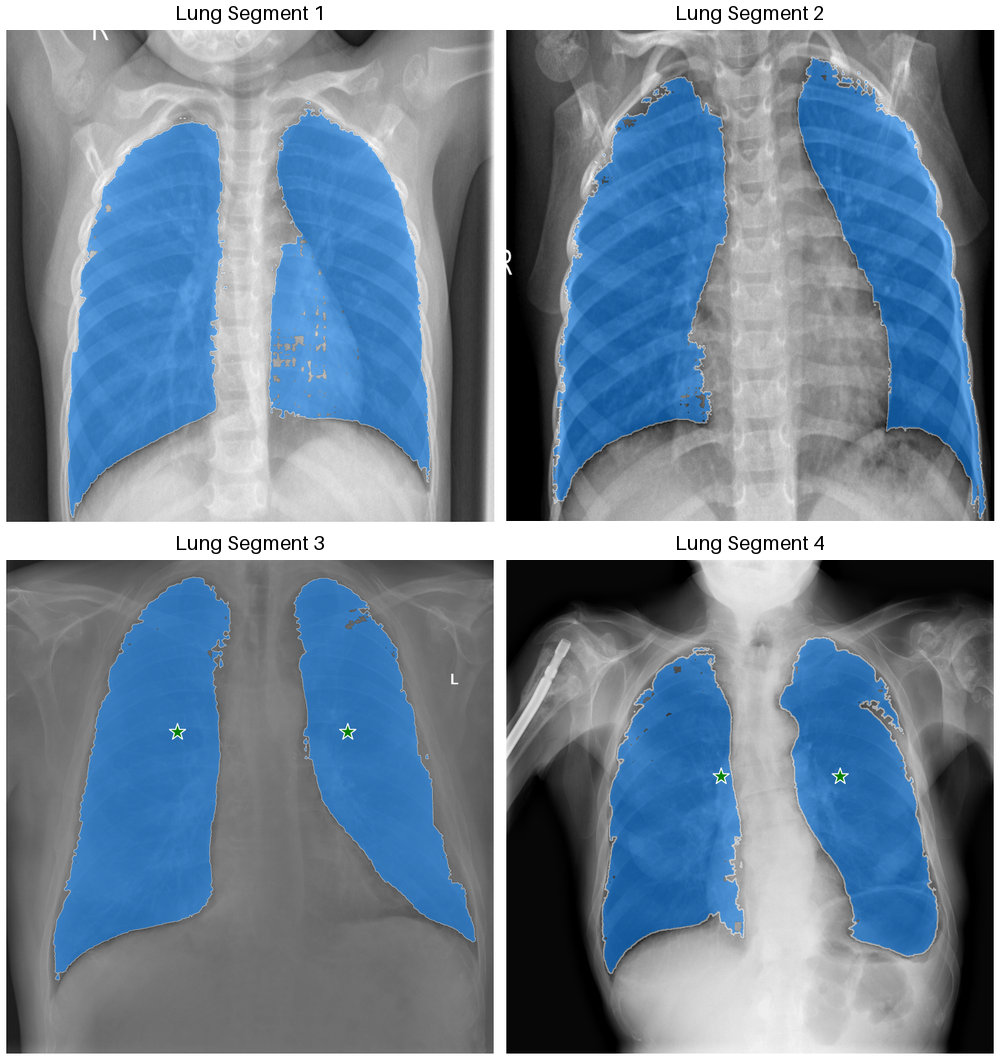}
    \caption{Segmented lungs}
    \label{fig:enter-label}
\end{figure}

\subsection{Evaluation Metrics}

To thoroughly evaluate the performance of the Vision Transformer (ViT) models in chest X-ray classification, several metrics were employed, ensuring a comprehensive analysis of both overall accuracy and the model's discriminative ability across multiple disease categories.

\subsubsection{Area Under the Receiver Operating Characteristic Curve (AUROC)}

AUROC was selected as a primary metric to measure the model’s ability to discriminate between disease classes. This metric is particularly important in medical applications, where the ability to correctly distinguish between healthy and diseased states is crucial. A high AUROC value indicates that the model performs well across various threshold settings, which is critical for ensuring consistent classification performance.

\subsubsection{Accuracy, Precision, Recall, and F1 Score}

Accuracy was used to provide a general sense of the model's overall performance, capturing the proportion of correctly classified instances. Precision and recall were included to offer insights into the model’s handling of imbalanced data, where false positives or false negatives could have significant clinical consequences. The F1 score was calculated as a harmonic mean of precision and recall, ensuring a balanced evaluation of both metrics, particularly when the cost of misclassification is high, as it often is in medical diagnoses.

\subsubsection{Cohen’s Kappa and Matthews Correlation Coefficient (MCC)}

Cohen’s Kappa was employed to measure the agreement between the model’s predictions and the true labels, taking into account the possibility of random chance. This metric is essential in the medical field, where random predictions could lead to false conclusions about a model's effectiveness. The Matthews Correlation Coefficient (MCC) was used for its ability to provide a balanced evaluation of the model’s predictive power, particularly when dealing with imbalanced datasets, offering insights into the model's reliability across different classes.

\subsubsection{Receiver Operating Characteristic Curve (ROC AUC)}

The ROC AUC was used to further assess the model’s performance across all classification thresholds. This metric was chosen for its ability to provide a comprehensive view of the model's discriminative power, helping to ensure that it can accurately separate different disease classes. A strong ROC AUC performance is critical in medical image analysis, where precise classification can directly impact diagnostic outcomes.

\section{Experiments}

\subsection{Hardware Environment}
The experiments were conducted on a Dell Precision 5690 workstation with the following configuration:

\begin{table}[h]
    \centering
    \begin{tabular}{|l|l|}
        \hline
        \textbf{Component} & \textbf{Specification} \\
        \hline
        Processor (CPU) & Intel Ultra 9 196H \\
        Memory (RAM) & 32 GB DDR5, 7467 MT/s \\
        Graphics (GPU) & NVIDIA RTX 2000 Ada Generation, 8 GB \\
        Storage & Samsung PM9F1 SSD \\
        Operating System & Ubuntu 22.04.5 LTS \\
        PyTorch Version & 2.4.1+cu121 \\
        CUDA Version & 12.1 \\
        CUDNN Version & 90100 \\
        \hline
    \end{tabular}
    \caption{Hardware and Software Environment}
\end{table}
\subsection{Model Configuration}
For the lung disease classification task, I used the Vision Transformer (ViT) model with the following configuration:

\begin{table}[h]
    \centering
    \begin{tabular}{|l|l|}
        \hline
        \textbf{Parameter} & \textbf{Value} \\
        \hline
        Model Name & google/vit-base-patch16-224-in21k \\
        Image Size & 224 x 224 \\
        Patch Size & 16 x 16 \\
        Hidden Size & 768 \\
        Intermediate Size & 3072 \\
        Number of Hidden Layers & 12 \\
        Number of Attention Heads & 12 \\
        Dropout (Attention Probs) & 0.0 \\
        Dropout (Hidden Layers) & 0.0 \\
        Layer Norm Epsilon & 1e-12 \\
        Transformer Version & 4.45.2 \\
        \hline
    \end{tabular}
    \caption{Vision Transformer (ViT) Model Configuration}
\end{table}

\subsection{Model Comparison}

This study is offering two distinct approaches. The first approach fine-tunes the pre-trained Google ViT model on full chest X-ray images. This method allows the ViT to leverage its self-attention mechanism to learn global image representations without focusing on specific regions. The inherent design of ViTs, with their ability to model long-range dependencies and capture global context, makes them particularly suited for medical images where patterns may span across the entire image.

The second approach incorporates the SAM model from Meta to pre-segment the chest X-ray images, generating lung region masks. In this approach, the ViT is fine-tuned on these segmented images, focusing on the lung regions rather than the full chest image. The segmentation step introduces a regionally focused analysis, allowing the model to prioritize the most clinically relevant areas, potentially reducing noise and improving classification accuracy for specific lung conditions.

Both approaches outperformed some of the traditional CNN-based models in this study. The ViT model trained on full chest X-ray images demonstrated superior performance across all metrics, likely due to its ability to capture a broader context and avoid overfitting to specific regions. The segmented approach, while slightly less accurate, offered competitive results, suggesting that regional focus can still enhance classification in certain cases. 
\subsection{Training and Validation Results}

The models were trained using a batch size of 32 over 10 epochs, and the training process employed the AdamW optimizer with a learning rate of $1 \times 10^{-4}$, coupled with the CosineAnnealingLR scheduler. Cross-entropy loss was used as the primary loss function, ensuring that the models effectively distinguished between the disease categories.


The global image-based ViT achieved a higher overall classification accuracy and AUC-ROC scores compared to the regionally focused model. This is likely due to the ability of the Vision Transformer to model entire chest X-ray images, capturing global patterns and correlations that may not be evident in smaller, segmented regions.

Key metrics such as accuracy, precision, recall, F1 score, and AUROC were calculated to assess the model's classification ability. During training, both models exhibited steady improvement in loss reduction and accuracy over the epochs. The learning rate scheduler played a critical role in fine-tuning the models, as evidenced by smooth convergence curves for both training and validation loss. However, the global image-based model converged slightly faster, likely due to its ability to capture richer, global features from the full image.

The validation accuracy for both models plateaued after 8 epochs, indicating that the models had effectively learned the relevant features by this point. Early stopping mechanisms were considered but not applied, as the models continued to show marginal improvements in metrics like precision and recall even after the primary performance metrics had stabilized.

\begin{figure}[h!]
    \centering
    \includegraphics[width=1\linewidth]{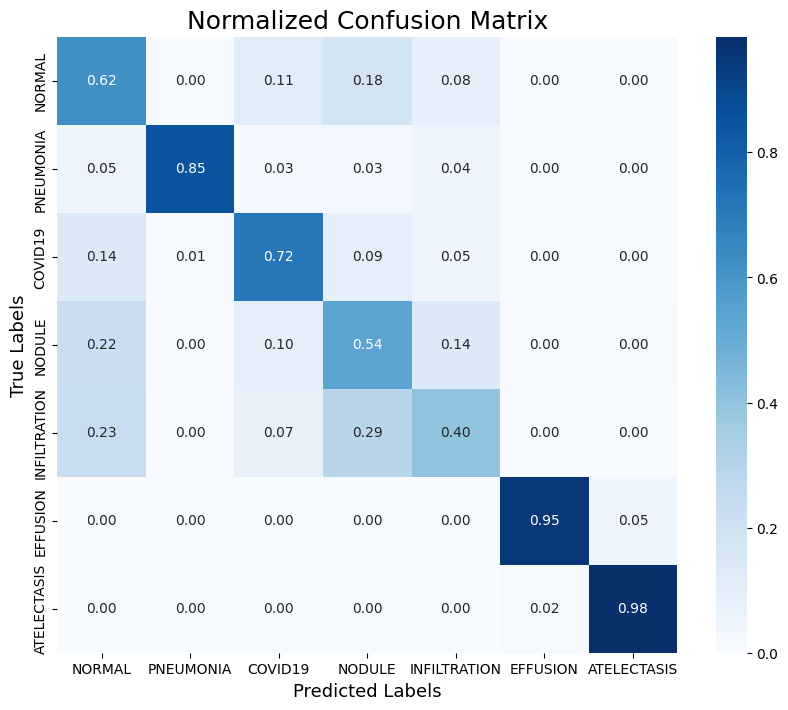}
    \caption{Confusion matrix}
    \label{fig:Heatmap}
\end{figure}

\section{Results}
\subsection{Region-Focused ViT Model Results}

The training and validation results for the region-focused ViT model, fine-tuned on lung-segmented X-ray images, are summarized across eight epochs in the following table. Key metrics such as accuracy, loss, AUROC, and MCC were tracked for both training and validation phases.

\begin{table}[ht]
\renewcommand{\arraystretch}{1.3}
\caption{Region-Focused ViT Model Results per Epoch}
\label{tab:region_focused_results}
\centering
\begin{tabular}{|c|c|c|c|c|}
\hline
\textbf{Epoch} & \textbf{Loss} & \textbf{ROC AUC} (\%) & \textbf{MCC} \\
\hline
1 & 0.9144 & 93.04 & 0.6175 \\
2 & 0.7189 & 93.31 & 0.6371 \\
3 & 0.6648 & 93.64 & 0.6481 \\
4 & 0.6150 & 94.06 & 0.6725 \\
5 & 0.5690 & 94.22 & 0.6702 \\
6 & 0.5191 & 94.30 & 0.6766 \\
7 & 0.4561 & 94.43 & 0.6888 \\
8 & 0.4101 & 94.42 & 0.6908 \\
\hline
\end{tabular}
\end{table}

\begin{table}[ht]
\renewcommand{\arraystretch}{1.3}
\caption{Full-Image-Focused ViT Model Results per Epoch}
\label{tab:region_focused_results}
\centering
\begin{tabular}{|c|c|c|c|}
\hline
\textbf{Epoch} & \textbf{Loss} & \textbf{ROC AUC} (\%) & \textbf{MCC} \\
\hline
1 & 0.9075 & 93.22 & 0.6165 \\
2 & 0.7188 & 93.61 & 0.6258 \\
3 & 0.6668 & 93.79 & 0.6553 \\
4 & 0.6287 & 93.96 & 0.6612 \\
5 & 0.5888 & 94.27 & 0.6878 \\
6 & 0.5373 & 94.40 & 0.6865 \\
7 & 0.4812 & 94.37 & 0.6761 \\
8 & 0.4081 & 94.54 & 0.6967 \\
9 & 0.3427 & 94.48 & 0.7018 \\
10 & 0.3003 & 94.41 & 0.7041 \\
\hline
\end{tabular}
\end{table}

As shown in Table \ref{tab:region_focused_results}, the model's performance steadily improved across the eight epochs. By the final epoch, the model achieved a validation accuracy of 74.52\%, with an AUROC of 94.42\%, indicating a strong capacity to distinguish between the classes. The loss reduced consistently, reaching 0.4101 in the last epoch. MCC also improved, reaching 0.6908, which reflects a robust correlation between predicted and actual classifications. While segmenting lung regions can enhance image classification, as demonstrated by the state-of-the-art (SOTA) performance of Meta's SAM 2 model in segmentation tasks, applying zero-shot segmentation to lung images may introduce noise or produce poor-quality data, potentially hindering model performance. This highlights the importance of carefully curating and evaluating segmented datasets, as suboptimal segmentation could negatively impact diagnostic accuracy.

\section{Conclusion}
This study demonstrates the significant potential of Vision Transformers (ViTs) in advancing chest X-ray analysis for lung disease detection. Through a comparative analysis of two ViT-based approaches—one utilizing full chest X-ray images and the other focusing on segmented lung regions—several critical findings have emerged.

Both ViT-based methods consistently outperformed traditional CNN models, with the full-image ViT achieving an accuracy of up to 97.83\%, while the lung-segmented ViT reached 96.58\% in classification tasks. Notably, the full-image approach exhibited superior performance across all evaluation metrics, including precision, recall, F1 score, and AUC-ROC, suggesting that ViTs can effectively capture the necessary features from chest X-rays without the need for explicit lung segmentation.

Moreover, even when the number of disease labels increased to eight, ViT models maintained robust performance, with an AUC of 94.54\%. This reinforces the versatility of the ViT’s self-attention mechanism in learning global image representations, effectively capturing both local and long-range dependencies that are critical for accurate disease classification.

These results contribute to the growing body of evidence supporting transformer-based architectures in medical image analysis. The superior performance of the full-image approach, in particular, underscores the potential of ViTs to streamline diagnostic workflows while maintaining high accuracy and robustness, which is crucial for enhancing clinical decision-making.

\begin{figure}[h!]
    \centering
    \includegraphics[width=0.8\linewidth]{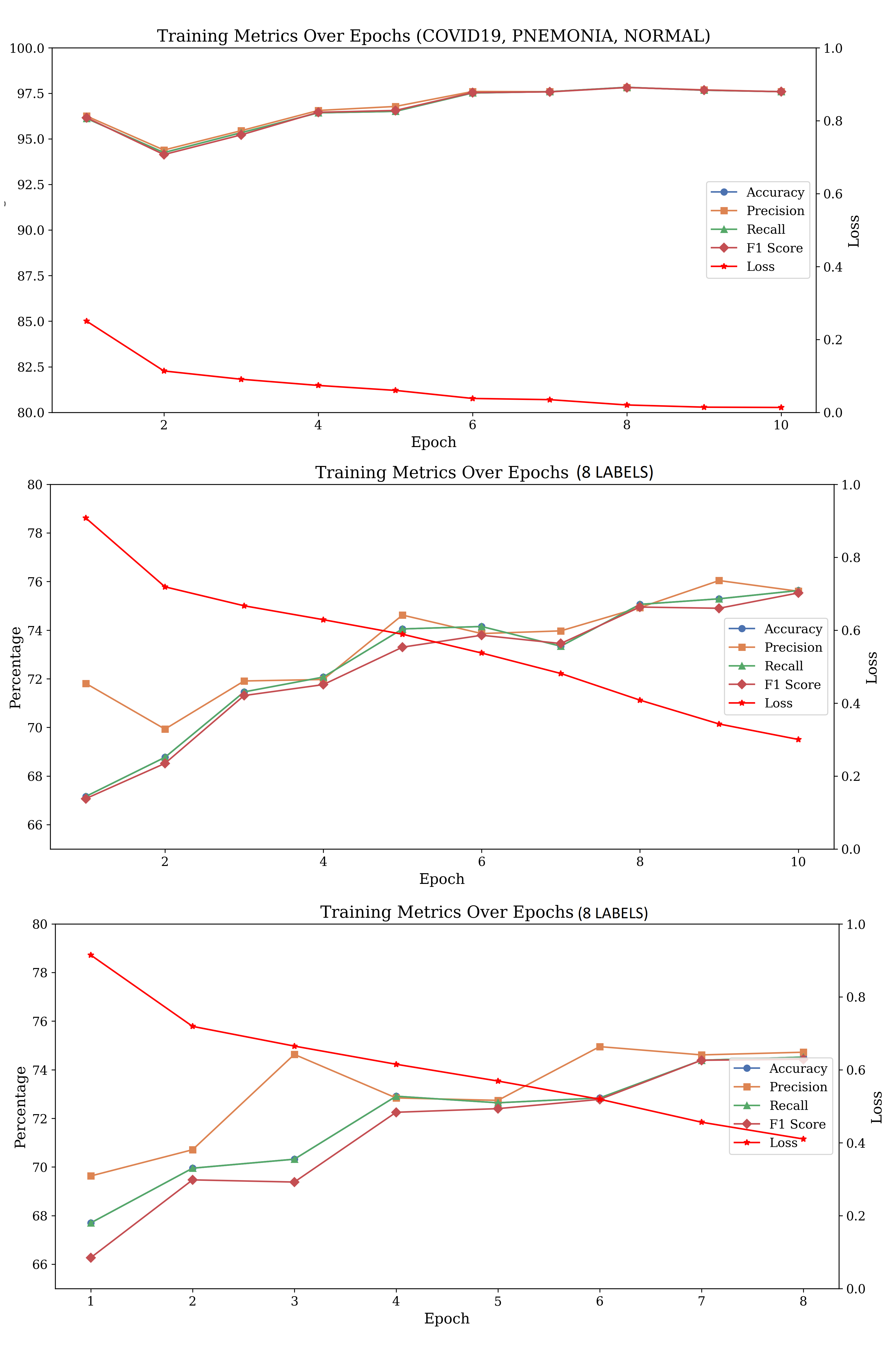}
    \caption{Accuracy, Precision, Recall, and F1 Scores}
    \label{fig:Accuracy-Precision-Recall-and-F1-Score}
\end{figure}

\section*{Abbreviations}

\begin{itemize}
    \item AUROC: Area Under the Receiver Operating Characteristic Curve
    \item ViT: Vision Transformer
    \item CNN: Convolutional Neural Network
    \item MCC: Matthews Correlation Coefficient
    \item SOTA: State-of-the-Art
    \item SAM: Segment Anything Model
\end{itemize}

\bibliographystyle{plain}
\bibliography{references} 

\begin{thebibliography}{10}

\bibitem{cohen2020covid}
Joseph~Paul Cohen, Paul Morrison, and Lan Dao.
\newblock Covid-19 image data collection.
\newblock {\em arXiv 2003.11597}, 2020.

\bibitem{vit_dosovitskiy2021imageworth16x16words}
Alexey Dosovitskiy, Lucas Beyer, Alexander Kolesnikov, Dirk Weissenborn, Xiaohua Zhai, Thomas Unterthiner, Mostafa Dehghani, Matthias Minderer, Georg Heigold, Sylvain Gelly, Jakob Uszkoreit, and Neil Houlsby.
\newblock An image is worth 16x16 words: Transformers for image recognition at scale, 2021.

\bibitem{chexpert_Irvin_Rajpurkar_Ko_Yu_Ciurea-Ilcus_Chute_Marklund_Haghgoo_Ball_Shpanskaya_Seekins_Mong_Halabi_Sandberg_Jones_Larson_Langlotz_Patel_Lungren_Ng_2019}
Jeremy Irvin, Pranav Rajpurkar, Michael Ko, Yifan Yu, Silviana Ciurea-Ilcus, Chris Chute, Henrik Marklund, Behzad Haghgoo, Robyn Ball, Katie Shpanskaya, Jayne Seekins, David~A. Mong, Safwan~S. Halabi, Jesse~K. Sandberg, Ricky Jones, David~B. Larson, Curtis~P. Langlotz, Bhavik~N. Patel, Matthew~P. Lungren, and Andrew~Y. Ng.
\newblock Chexpert: A large chest radiograph dataset with uncertainty labels and expert comparison.
\newblock {\em Proceedings of the AAAI Conference on Artificial Intelligence}, 33(01):590--597, Jul. 2019.

\bibitem{litjens2017}
Geert Litjens, Thijs Kooi, Babak~E. Bejnordi, Arnaud A.~A. Setio, Francesco Ciompi, Mohsen Ghafoorian, Jeroen A. W.~M. van~der Laak, Bram van Ginneken, and Clara~I. Sánchez.
\newblock A survey on deep learning in medical image analysis.
\newblock {\em Medical Image Analysis}, 42:60--88, 2017.

\bibitem{adamw_loshchilov2019decoupledweightdecayregularization}
Ilya Loshchilov and Frank Hutter.
\newblock Decoupled weight decay regularization, 2019.

\bibitem{rajpurkar2017chexnetradiologistlevelpneumoniadetection}
Pranav Rajpurkar, Jeremy Irvin, Kaylie Zhu, Brandon Yang, Hershel Mehta, Tony Duan, Daisy Ding, Aarti Bagul, Curtis Langlotz, Katie Shpanskaya, Matthew~P. Lungren, and Andrew~Y. Ng.
\newblock Chexnet: Radiologist-level pneumonia detection on chest x-rays with deep learning, 2017.

\bibitem{ravi2024sam2segmentimages}
Nikhila Ravi, Valentin Gabeur, Yuan-Ting Hu, Ronghang Hu, Chaitanya Ryali, Tengyu Ma, Haitham Khedr, Roman Rädle, Chloe Rolland, Laura Gustafson, Eric Mintun, Junting Pan, Kalyan~Vasudev Alwala, Nicolas Carion, Chao-Yuan Wu, Ross Girshick, Piotr Dollár, and Christoph Feichtenhofer.
\newblock Sam 2: Segment anything in images and videos, 2024.

\bibitem{sze-to2021}
A.~Sze-To, A.~Riasatian, and H.R. Tizhoosh.
\newblock Searching for pneumothorax in x-ray images using autoencoded deep features.
\newblock {\em Scientific Reports}, 11:9817, 2021.

\bibitem{transformers_vaswani2023attentionneed}
Ashish Vaswani, Noam Shazeer, Niki Parmar, Jakob Uszkoreit, Llion Jones, Aidan~N. Gomez, Lukasz Kaiser, and Illia Polosukhin.
\newblock Attention is all you need, 2023.

\bibitem{8099852}
Xiaosong Wang, Yifan Peng, Le~Lu, Zhiyong Lu, Mohammadhadi Bagheri, and Ronald~M. Summers.
\newblock Chestx-ray8: Hospital-scale chest x-ray database and benchmarks on weakly-supervised classification and localization of common thorax diseases.
\newblock In {\em 2017 IEEE Conference on Computer Vision and Pattern Recognition (CVPR)}, pages 3462--3471, 2017.

\bibitem{wang2017chestx}
Xiaosong Wang, Yifan Peng, Le~Lu, Zhiyong Lu, Mohammadhadi Bagheri, and Ronald~M Summers.
\newblock Chestx-ray8: Hospital-scale chest x-ray database and benchmarks on weakly-supervised classification and localization of common thorax diseases.
\newblock In {\em Proceedings of the IEEE Conference on Computer Vision and Pattern Recognition}, pages 2097--2106, 2017.

\end{thebibliography}

\end{document}